\documentclass[prl,aps,amssymb,twocolumn,showpacs,superscriptaddress]{revtex4}
\usepackage{graphicx}

\begin{document}

\title{Dynamics of Quantum Noise in a Tunnel Junction under ac Excitation}

\author{J.~Gabelli}
\author{B.~Reulet}
\affiliation{Laboratoire de Physique des Solides, UMR8502, b\^atiment 510, Universit\'e Paris-Sud, 91405 ORSAY Cedex, France}

\date{\today}
\begin{abstract}
We report the first measurement of the \emph{dynamical response} of shot noise (measured at frequency $\omega$) of a tunnel junction to an ac excitation at frequency $\omega_0$. The experiment is performed in the quantum
regime, $\hbar\omega\sim\hbar\omega_0\gg k_BT$ at very low
temperature $T=35$mK and high frequency $\omega_0/2\pi=6.2$ GHz. We observe that the noise responds in phase with the excitation, but not adiabatically. The results are in very good agreement with a prediction based on a new current-current correlator.
\end{abstract}
 \maketitle

Physics of current fluctuations in mesoscopic conductors has drawn a
lot of interest during the last 15 years, first as a challenge to
our understanding of the full statistics of charge transport, and
second with the hope to use fluctuations as a tool to probe
conduction mechanisms in complex systems. A simple characterization
of current fluctuations is given by their spectral density
$S_2(\omega)$ measured at a frequency $\omega$: $S_2(\omega)=\langle
i(\omega)i(-\omega)\rangle$. Here $i(\omega)$ is the Fourier
component of the classical, fluctuating current at frequency
$\omega$, generally measured after amplification. Noise at low
frequency has been addressed in numerous systems, and some finite
frequency measurements in the simplest systems have been reported
(for a review, see ref. \cite{BuBlan}). The remarkable agreement
between experiments and theory bears witness of the deep
understanding of shot noise that has been achieved.

Noise measurements have also proven to be a useful tool to probe conduction mechanisms in certain situations. In particular, the evidence for a non-integer effective charge has demonstrated the effects of interations (like in the fractional quantum Hall effect \cite{FQHE}) and interferences (like in the Andreev interferometer \cite{AndreevInterf}), hallmarks of the quantum nature of transport. However, little has been obtained on the \textit{dynamics} of transport. For example,  measuring the frequency dependence of conductance or noise of a  diffusive wire does not provide the simplest -- though essential -- parameter of electron motion in the sample, the diffusion time \cite{Hekking}. Such an information can be obtained only from the frequency dependence of the quantum corrections to conductance \cite{Pieper}, that  of the third cumulant of current fluctuations \cite{Pilgram}, or that of the recently introduced noise thermal impedance (NTI) \cite{NTI}.
The NTI characterizes the \emph{dynamical response of the noise spectral density} $S_2(\omega)$ to an ac+dc bias $V(t)=V+\delta V\cos\omega_0t$. It measures how, in phase and in quadrature with the excitation, the amplitude of the noise is modulated by the ac voltage, in the same way as the ac conductance $G(\omega_0)$ measures the modulation of the average current, $\delta I=G(\omega_0)\delta V$. The dependence of the NTI on the excitation frequency $\omega_0$ unveils information that ac conductance and noise at finite frequency $S_2(\omega)$ do not provide.

However, the NTI is well defined only if the time scale $\tau$ over which noise is detected fulfills two conditions: 1) it is much longer than the typical period of the fluctuating current, i.e. $\tau\gg\omega^{-1}$; 2) it is much shorter than the period of the ac excitation, i.e. $\tau\ll\omega_0^{-1}$.
This picture, which mixes time and frequency, is conceptually restricted to $\omega_0\ll\omega$. It is the goal of
the present letter to extend the notion of noise dynamics to arbitrary frequencies $\omega$ and $\omega_0$, and to provide a measurement in a regime clearly beyond that of the NTI.

The paper is organized as follows: 1) we determine that the correct current correlator which describes our noise dynamics at high frequency is $\langle i(\omega)i(\omega_0-\omega)\rangle$. 2) We design an experimental setup to measure this correlator in the quantum regime  $\hbar\omega\sim\hbar\omega_0\gg k_BT$. 3) We report the first measurement of the dynamics of quantum noise in a tunnel junction, the simplest coherent conductor, in a regime far beyond what could have been expected from low frequency considerations. We observe that the noise of the tunnel junction responds in phase with the ac excitation, but its response is not adiabatic, as obtained in the limit of slow excitation. Our data are in quantitative agreement with a calculation we have performed, the details of which are not included in this letter \cite{JGBRth}. We also report the first measurement of photo-assisted finite frequency noise, i.e. the effect of an ac excitation on the \emph{time averaged} noise $S_2(\omega)$ for $\hbar\omega\gg k_BT$.

\emph{Noise and photo-assisted noise} ---
In order to introduce the correlator that describes our noise dynamics, we start with those which describe noise and photo-assisted noise. The spectral density of the current fluctuations at frequency $\omega$ of a coherent conductor with no internal dynamics biased by a dc voltage $V$ is given by \cite{BuBlan} :
\begin{equation}
S_2(V,\omega)=(F/2) \left[ S_2^0( \omega_+)+S_2^0 ( \omega_-)\right] +(1-F)S_2^0(\omega)
\label{eqSvsS0}
\end{equation}
where $F$ is the Fano factor and $\omega_\pm=\omega\pm eV/\hbar$. $S_2^0(\omega)$ is the Johnson-Nyquist equilibrium noise, $S_2^0(\omega)=2G\hbar\omega\coth[\hbar\omega/(2k_BT)]$ and $G$ is the conductance. This simple relation comes from the bias voltage affecting the electrons wavefunctions only through a phase factor $\exp (ieVt/\hbar)$ \cite{Tien}. For a tunnel junction $F\sim1$ and the last term of Eq. (\ref{eqSvsS0}) vanishes. At low temperature, the $S_2$ vs. $V$ curve has kinks at $eV=\pm\hbar\omega$, as clearly demonstrated in our measurement, see fig. \ref{fig:Si} top.
When an ac bias voltage $\delta V\cos\omega_0t$ is superimposed on the dc one, the electrons wavefunctions acquire an extra phase factor $\sum_n J_n(z)\exp (in\omega_0t)$ where $J_n$ is the ordinary Bessel function and $z=e\delta V/(\hbar\omega_0)$.  The noise  at frequency $\omega$ is modified by the ac bias, to give \cite{LevitovACAB}:
\begin{equation}
S_2^{pa}(V,\omega)=\sum_{n=-\infty}^{+\infty}J_n^2(z) S_2(V-n\hbar\omega_0/e,\omega)
\label{eqSpa}
\end{equation}
This effect, called \emph{photo-assisted noise}, has been measured for $\omega=0$ \cite{PAT_Rob}. We show below the first measurement of photo-assisted noise at finite frequency $\omega$. The multiple steps separated by $eV=\hbar\omega_0$ are well pronounced, see fig. \ref{fig:Si} bottom. Let us emphasize that it corresponds to noise \emph{averaged over time}.

\emph{From noise to noise dynamics} ---
The noise spectral density is mathematically well defined as the one frequency current-current correlator $S_2(\omega)=\langle i(\omega)i(-\omega)\rangle$ \cite{footsym}. In contrast, the NTI is well defined in terms of time-dependent noise temperature \cite{NTI} but lacks a mathematical definition in terms of current-current correlator, which is essential if we want to extend it to arbitrary frequencies. In order to infer such a definition, we recall how noise at high frequency is measured. Experimentally, $S_2(\omega)$ is obtained by 1) filtering the current to keep only the $\omega$ Fourier component, to give $I_\omega(t)=i(\omega)e^{i\omega t}+i(-\omega)e^{-i\omega t}$; 2) taking the square of the filtered current and average it over time, $\langle I_\omega^2(t)\rangle=2S_2(\omega)$. In this procedure, the current at frequency $\omega$ beats with itself to give a dc quantity. In the presence of a slow varying ac voltage, we expect the noise measured with this method to oscillate in phase with the excitation at  frequency $\omega_0$. Such a signal comes from the beating of two Fourier components of the current separated by $\pm\omega_0$, i.e. is related to the correlator $\langle i(\omega) i(\omega_0-\omega)\rangle$. This correlator is the simplest one that expresses the dynamical response of noise.

\begin{figure}
\includegraphics[width= 0.8\columnwidth]{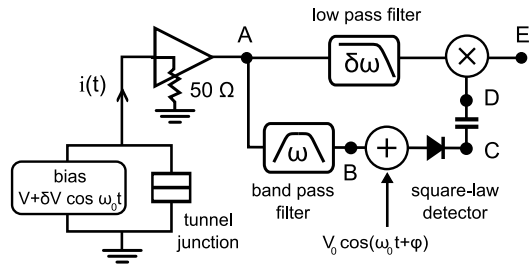}
\caption{ Experimental setup for the measurement of the noise dynamics $X(\omega_0,\omega)$ for $\omega \sim \omega_0$. The symbol $\oplus$ represents a combiner, which output is the sum of its two inputs. The symbol $\otimes$ represents a multiplier, which output is the product of its two inputs. The diode symbol represents a square law detector, which output is proportional to the low frequency part of the square of its input.}
\label{fig:schem_X}
\end{figure}

\emph{Principles of detection} ---
The setup we have built to measure the correlator $\langle i(\omega) i(\omega_0-\omega)\rangle$ is depicted on Fig. \ref{fig:schem_X}. Current
is measured as a voltage drop across a load resistor (the $50\Omega$ input impedance of the amplifier) that is amplified to give the signal at point A. In the lower arm, a bandpass filter selects Fourier components of the current around $\pm\omega$ within $\pm\delta\omega$ (we choose $\omega, \delta\omega>0$). After the filter, the voltage is:
$$
v_B(t)\propto\int_{\omega-\delta\omega}^{\omega+\delta\omega} d\omega_1 [i(\omega_1)e^{i\omega_1t} + i(-\omega_1)e^{-i\omega_1t}]
$$
We then add to $v_B$ a reference voltage $V_0\cos(\omega_0t+\varphi)$ at the same frequency as the excitation of the sample. A square-law detector (i.e., a power meter), represented by a diode symbol on Fig. \ref{fig:schem_X}, takes the square of this voltage and cuts-off high frequencies. At point C there is a dc voltage $\langle v_C\rangle$ which has three contributions: 1) the reference beating with itself, $\propto V_0^2/2$; 2) the reference beating with $i(\omega_0)$, $\propto V_0\delta V \mathrm{Re}\, G(\omega_0)$ with $G(\omega_0)$ the (complex) conductance of the sample at frequency $\omega_0$ ; 3) each Fourier component of the current beating with itself to give the noise integrated over the band of the filter, $\propto S_2(\omega)\delta\omega$. These dc contributions are removed by a capacitor. The voltage at point D, $v_D(t)=v_C(t)-\langle v_C\rangle$ is the result of the beating of the reference with the current $i(\omega)$ for $|\omega-\omega_0|<\delta\omega$ and $\omega\neq\omega_0$:
$$
v_D(t)\propto V_0 \int_{\omega-\delta\omega}^{\omega+\delta\omega} d\omega_1 [i(\omega_1)e^{i(\omega_1-\omega_0)t} + i(-\omega_1)e^{-i(\omega_1-\omega_0)t}]
$$
plus terms involving the square of the current, $i(\omega_1)i(-\omega_2)$ which are much smaller since $V_0$ is large \cite{noteS3}. $v_D(t)$ is multiplied with the low frequency part of $i(t)$ (upper arm of the setup). The dc output $\langle v_{E}\rangle$ of the multiplier is, for $\varphi=0$, proportional to the quantity:
\begin{equation}
X(\omega_0,\omega)=\frac{1}{2} \, \left\{
\langle i(\omega)i(\omega_0-\omega\rangle + \langle i(-\omega)i(\omega-\omega_0)\rangle \right\}
\end{equation}
Our setup is restricted to $\omega\sim\omega_0$ due to  the small output bandwidth of the power detector. However,  this restriction corresponds precisely to the most interesting situation \cite{JGBRth,Kinder}. Note that $\omega$ can be as large as wanted, in particular we achieve the quantum regime $\hbar\omega\sim\hbar\omega_0\gg eV,k_BT$.

\emph{Experimental setup} --- In order to demonstrate the relevance of the noise dynamics, we have chosen to perform the measurement on the simplest system that exhibits well understood shot noise, the tunnel junction. The sample is an Al/Al oxide/Al tunnel junction similar to that used for noise thermometry \cite{Lafe}. We apply a 0.1 T perpendicular magnetic field to turn the Al normal. The junction is mounted on a rf sample holder placed on the mixing chamber of a dilution refrigerator. The sample is dc voltage biased, ac biased at $\omega_0/2\pi=6.2$ GHz, and ac coupled to a microwave 0.01-8 GHz cryogenic amplifier. To preselect $\omega$ we use a $5.7-6.7$ GHz band-pass filter (Fig. \ref{fig:schem_X}, lower arm). The low frequency part of the current, at frequency $\omega-\omega_0$, is selected by a $200$ MHz low pass filter (Fig. \ref{fig:schem_X}, upper arm). The power detector has an output bandwidth of $\delta\omega/2\pi\sim 200$ MHz, which limits the frequencies $\omega$ contributing to the signal: $|\omega|\in[\omega_0-\delta\omega,\omega_0+\delta\omega]$.   The resistance of the sample $R_0=44.2 \, \Omega$ is close to $50 \, \Omega$ to provide a good matching to the coaxial cable and avoid reflection of the ac excitation.

\begin{figure}
\includegraphics[width= 0.8\columnwidth]{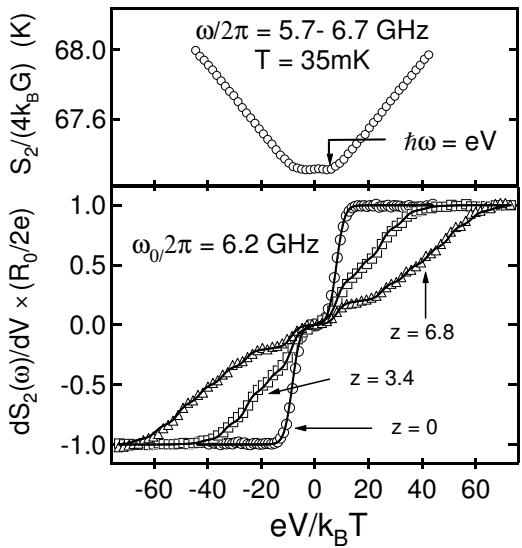}
\caption{Top: Measured noise temperature $T_N=S_2(\omega)/(4k_BG)$ of the sample plus the amplifier with no ac excitation. Bottom: measured differential noise spectral density $dS_2(\omega)/dV$ for various levels of excitation $z=e\delta V/(\hbar\omega_0)$. $z\neq 0$ corresponds to photo-assisted noise. Solid lines are fits with Eq. (\ref{eqSpa}).}
\label{fig:Si}
\end{figure}

\emph{Calibration} --- Several quantities need to be determined in order to make quantitative comparisons between experiment and theory: the electron temperature $T$, the high frequency ac voltage across the sample $\delta V$ (which we express through the dimensionless parameter $z=e\delta V/\hbar\omega_0$), the gain of the detection and the phase difference between the excitation and the detection at $\omega_0$ (in order not to mix in-phase and out-of-phase responses), which is tuned by $\varphi$, see Fig. \ref{fig:schem_X}. The temperature is determined by the measurement of $S_2(\omega)$ vs. dc voltage \textit{with no ac excitation} ($\delta V=V_0=0$), obtained at point $C$: $\langle v_{C}\rangle\propto S_2(\omega)$, see Fig. \ref{fig:Si} top. We find $T=35$mK, i.e. higher than the phonon temperature (15mK), probably due to the broadband emissions from the amplifier towards the sample. We conclude that $\hbar\omega/k_BT\sim8.5$. The ac voltage $\delta V$ is deduced from photo-assisted noise measurement, i.e. measurement of $S_2(\omega)$ (averaged over time) in the presence of an excitation at frequency $\omega_0$, see Fig. \ref{fig:Si} bottom \cite{notedS2}. Note that we measure the photo-assisted noise in the regime $\omega\sim\omega_0$ and not at low frequency $\omega\ll\omega_0$, as in ref. \cite{PAT_Rob}. Such a measurement had never been reported before. Since the tunnel junction is not very sensitive to heating due to external noise, we clearly identify the plateaus due to multi-photons transitions, hardly visible in ref. \cite{PAT_Rob}. The fit of the bottom curves of fig. \ref{fig:Si} with Eq. (\ref{eqSpa}) provides the value of $\delta V$. The gain of the detection has been calibrated for each component at room temperature, and from the measurement and fit of $S_2(\omega)$ vs. $V$ for $\omega/2\pi\sim6.2$ GHz and $\omega/2\pi\sim100$ MHz. We obtain an agreement with the theory (see below) with a precision of $\sim20$\%. We discuss the phase $\varphi$ later.

\begin{figure}
\includegraphics[width= 0.9\columnwidth]{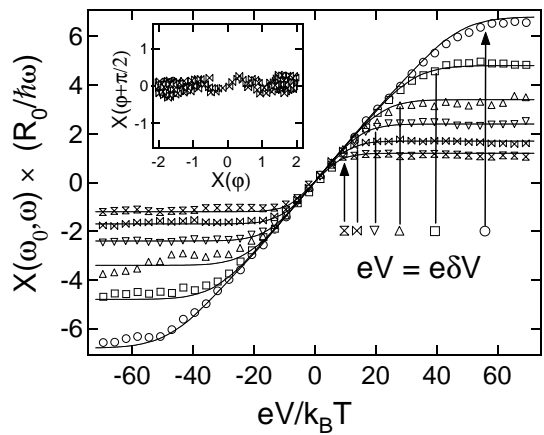}
\caption{Normalized noise dynamics $X(\omega_0,\omega)$ vs. normalized dc bias for various excitation levels $\delta V$. Symbols are data; solid lines are fits with Eq. (\ref{eq:X}) for $\omega=\omega_0$. Vertical arrows correspond to the cross-over $e\delta V=eV$. \textit{Inset:} Nyquist representation of $X(\omega_0,\omega)$ for $z=1.7$ (in arbitrary units). The in-phase and out-of-phase responses are measured by shifting the phase $\varphi$ of the reference signal by $90 ^{\circ}$.}
\label{fig:X}
\end{figure}

\begin{figure}
\includegraphics[width= 0.9\columnwidth]{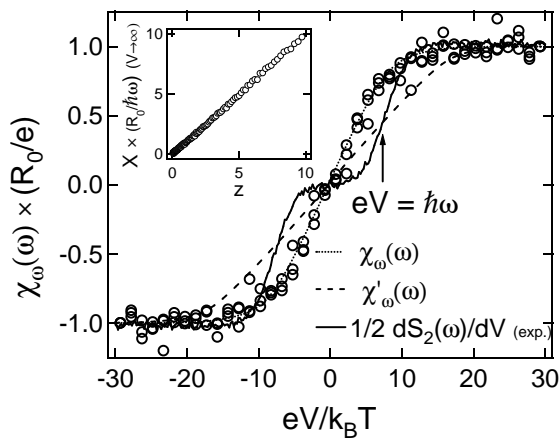}
\caption{Normalized noise susceptibility $\chi_{\omega}(\omega)$ vs. normalized dc bias. Symbols: data for various levels of excitation ($z = 0.85$,
$0.6$ and $0.42$). Dotted and dashed lines: fits of $\chi_{\omega}(\omega)$ (Eq. (\ref{eq:chi0})) and $\chi'_\omega(\omega)=[\chi_{\omega}(\omega)+\chi_{\omega}(-\omega)]/2$. Solid line: $(1/2)dS_2/dV$ (experimental), as a comparison. \textit{Inset:} Linear dependence of $X(\omega,\omega)$ vs. $z$ for large dc bias.}
\label{fig:chi}
\end{figure}

\emph{Experimental results} --- We have measured $X(\omega_0,\omega)$ vs. bias $V$ and ac excitation $\delta V$ for $\omega_0/2\pi=6.2$ GHz and $\omega/2\pi$ integrated over the band $6-6.4$ GHz. Fig. 3 shows data for various levels of ac excitation $z=e\delta V/(\hbar\omega_0)$ . In contrast with $dS_2(\omega)/dV$, $X(\omega_0,\omega)$ has no plateau for $eV<\hbar \omega$ but presents a cross-over at $eV\sim e\delta V$  for significant excitation $z>1$, see arrows on Fig. \ref{fig:X}. At high dc bias ($V\gg\delta V$), $X(\omega_0,\omega)$ is independent of $V$ but proportional to $\delta V$, see inset of Fig. \ref{fig:chi}. Using the techniques described in \cite{BuBlan}, we have calculated the correlator that corresponds to our experimental setup. We find for a tunnel junction \cite{JGBRth}:
\begin{equation}\label{eq:X}
\begin{array}{l}
X(\omega_0,\omega)  =  (1/2) \sum_n J_n(z)J_{n+1}(z)\times  \\
  \hspace{0.5cm} \left[ S_2^0(\omega_+ +n\hbar\omega_0)
   -S_2^0(\omega_- +n\hbar\omega_0) \right]  \\
\end{array}
\end{equation}
Note the similarity with the expression giving the photo-assisted noise, Eq. (\ref{eqSpa}). Note however that the sum in Eq. (\ref{eq:X}) expresses the \emph{interference} of the processes where $n$ photons are absorbed and $n\pm1$ emitted (or vice-versa), each absorption / emission process being weighted by an amplitude $J_n(z)J_{n\pm1}(z)$. As can be seen on Fig. \ref{fig:X}, our data is in quantitative agreement with the calculation with no fitting parameter. The theory predicts that, due to zero dwell time, there should be no out-of-phase response of noise for the tunnel junction, i.e. $X$ is real. We could not determine the absolute phase between the detected signal and the excitation voltage at the sample level. However we have varied the phase $\varphi$ to measure the two quadratures of the signal. We have always found that all the signal can be put on one quadrature only (independent of dc and ac bias, see inset of Fig. \ref{fig:X}), in agreement with the prediction.

 In order to emphasize the linear response regime of $X(\omega_0,\omega)$ to the excitation $\delta V$, we define the \emph{noise susceptibility} as:
$\chi_{\omega_0}(\omega)=\lim_{\delta V\rightarrow0} X(\omega_0,\omega)/\delta V$.
$\chi_{\omega_0}(\omega)$ expresses the effect, to first order in $\delta V$, of a small excitation at frequency $\omega_0$ to the noise measured at frequency $\omega$. We show on Fig. \ref{fig:chi} the data for $X(\omega_0,\omega)/\delta V$ at small injected powers as well as the theoretical curve for $\chi_{\omega_0}(\omega=\omega_0)$:
\begin{equation}
\chi_{\omega}(\omega)=\chi_{\omega}(0)= (1/2)(e/\hbar\omega)[S_2^0(\omega_+)-S_2^0(\omega_-)]
\label{eq:chi0}
\end{equation}
All the data fall on the same curve, as predicted, and are very well fitted by the theory. The cross-over occurs now for $eV\sim\hbar\omega$. However, $\chi_{\omega}(\omega)$ is clearly different from the adiabatic response of noise $dS_2(\omega)/dV$ (solid line on Fig. 4). This is the central result of our work.

In the limit $\delta V\rightarrow0$ and $\omega_0\rightarrow 0$ (with $z\ll1$), Eq. (\ref{eq:chi0}) reduces to
$\chi_\omega(0)\sim (1/2)(dS_2/dV)$. The factor $1/2$ comes from the
fact that the sum of frequencies, $\pm(\omega+\omega_0)$ (here $\sim12$ GHz), is not
detected in our setup. It is remarkable that Eq. (\ref{eq:X}), which fits very well our experimental data, is not invariant upon sign reversal of $\omega$ (or $\omega_0$). Another quantity can be formed, which restores this symmetry: $X'(\omega_0,\omega)= [X(\omega_0,\omega)+X(\omega_0,-\omega)]/2$. The associated susceptibility for $\omega_0=\omega$, $\chi'_{\omega}(\omega)$ is plotted on Fig. \ref{fig:chi}: it does not fit the data. It could be measured if the sum $\pm(\omega+\omega_0)$ was detected as well as the difference $\pm(\omega-\omega_0)$, i.e. with another experimental setup.

The noise susceptibility we have measured allows to characterize the dynamics of the fluctuations in a conductor. It is also a central quantity in the understanding of the environmental effects on high order cumulants of noise, in particular the third cumulant $S_3$ \cite{S3BR,Beenakker}. In this context, due to the finite impedance of the environment, the voltage fluctuations of the sample at frequency $\omega'$ are modulated by its noise at frequency $\omega$, which modifies $S_3(\omega,\omega')$. This contribution clearly involves the quantity $S_2(\omega)\chi_{\omega}(\omega')$ \cite{Kinder}. Thus the understanding of the noise susceptibility is crucial to future studies of higher order cumulants at finite frequency, in particular in the quantum regime.

We are very grateful to L. Spietz for providing us with the sample that he fabricated at Yale University.
We thank M. Aprili, H. Bouchiat, R. Deblock, M. Devoret, D. Est\`eve, G. Montambaux, F. Pierre, H. Pothier, J.-Y. Prieur, D.E. Prober, I. Safi and C. Urbina for fruitful discussions. This work was supported by ANR-05-NANO-039-02.



\begin{thebibliography}{99}
\bibitem{footsym} We do not discuss here the problem of ordering of the corresponding quantum current operators; in particular, we consider $S_2(\omega)=S_2(-\omega)$.
\bibitem{notedS2} To get rid of slow drifts of the amplifiers, we measure $dS_2(\omega)/dV$ with an additionnal 77 Hz, small voltage modulation and usual lock-in detection.
\end{thebibliography}
\end{document}